      [1999/12/01 v1.4c Il Nuovo Cimento]
\documentclass{cimento}

\input epsf
\usepackage{graphicx}   

\title{Highlights from the ARGO-YBJ experiment}
\author{G. Di Sciascio\from{ins:x}\ETC,   for the ARGO-YBJ Collaboration}
\instlist{\inst{ins:x} INFN, Sez. Roma Tor Vergata, Via della Ricerca
Scientifica 1, I-00133 Roma, Italy}
\begin{document}

\maketitle

\begin{abstract}
The ARGO-YBJ experiment is a multipurpose detector exploiting the full coverage approach at very high altitude. The apparatus, in stable data taking since November 2007 with an energy threshold of a few hundreds of GeV and a duty-cycle of about 90\%, is located at the YangBaJing Cosmic Ray Laboratory (Tibet, P.R. China, 4300 m a.s.l., 606 g/cm$^2$).
A number of interesting results are available in Cosmic Ray Physics and in Gamma Ray Astronomy after the first 3 years of stable data taking.
In this paper Gamma-Ray Astronomy results are summarized. 
\end{abstract}

\section{Introduction}

The ARGO-YBJ experiment is in stable data taking since November 2007 with a duty cycle $>$85\% and with excellent performance.
The large field of view ($\sim$2 sr) and the high duty cycle allow
a continuous monitoring of the sky in the declination band
from -10$^{\circ}$ to 70$^{\circ}$.
Several interesting results are available in cosmic ray (CR) physics as well as in gamma ray astronomy with the first 3 years of operation.

We observed TeV emission from 3 sources with a significance greater than 5 standard deviations (s.d.): Crab Nebula, Mrk421 and MGRO J1908+06.
In particular we observed the Crab Nebula with a significance of 14 s.d. in $\sim$800 days.
A detailed long-term monitoring of the Mrk421 flaring activity has been carried out in 2008-2009. A clear correlation with X-ray data was found. The relation between spectral index and flux has been studied in one single flare (June 2008) \cite{aielli10} and averaging the data over 2 years.

Working in scaler mode \cite{aielli08} ARGO-YBJ has performed a search for emission from GRBs in coincidence with 93 events observed by satellites (16 with known redshift), setting upper limits on the fluence between 10$^{-5}$ and 10$^{-3}$ erg cm$^{-2}$ in the 1 - 100 GeV energy range \cite{aielli09a}.

A medium-scale CR anisotropy has been observed with a significance greater than 10 s.d. at proton median energy of about 2 TeV. 
Two excesses (corresponding to a flux increase of $\sim$0.1\%) are observed by ARGO-YBJ around the positions $\alpha\sim$ 120$^{\circ}$, $\delta\sim$ 40$^{\circ}$ and $\alpha\sim$ 60$^{\circ}$, $\delta\sim$ -5$^{\circ}$ \cite{vernetto09}, in agreement with a similar detection reported by the Milagro Collaboration \cite{mil08}.
The origin of this medium-scale anisotropy is puzzling. In fact, these regions have been interpreted as excesses of hadronic CRs, but TeV CRs are expected to be randomized by the magnetic fields. Understanding these anisotropies should be a high priority as they are probably due to a nearby source of CRs, as suggested by some authors (see, e.g., \cite{markov}).

With 2 years of data we carried out a 2D measurement of the CR large-scale anisotropy to investigate detailed structural information beyond the simple Right Ascension profiles. 

The p-air cross section has been measured in the range 1 - 100 TeV and the corresponding p-p cross section inferred \cite{aielli09b}. 

The light-component (p+He) spectrum of primary CRs has been measured in the range 5 - 250 TeV. The preliminary results are in good agreement with the CREAM balloon data. For the first time direct and ground-based measurements overlap for a wide energy range thus making possible the cross-calibration of the different experimental techniques.

A measurement of the $\overline{p}/p$ ratio at few-TeV energies has been performed setting two upper limits at the 90\% confidence level: 5\% at 1.4 TeV and 6\% at 5 TeV \cite{cris10}.
In the few-TeV range these results are the lowest available, useful to constrain models for antiproton production in antimatter domains.

The main results after about 3 years of stable operation are summarized in \cite{vulcano10}.
In this paper gamma-ray astronomy results are summarized. 

\section{The ARGO-YBJ experiment}

The ARGO-YBJ experiment, located at the YangBaJing Cosmic Ray
Laboratory (Tibet, P.R. China, 4300 m a.s.l., 606 g/cm$^2$), is currently the only air shower array exploiting the full coverage approach at high
altitude, with the aim of studying the cosmic radiation at an
energy threshold of a few hundred GeV.

The detector is composed of a central carpet large $\sim$74$\times$
78 m$^2$, made of a single layer of Resistive Plate Chambers
(RPCs) with $\sim$93$\%$ of active area, enclosed by a guard ring
partially ($\sim$20$\%$) instrumented up to $\sim$100$\times$110
m$^2$. The apparatus has modular structure, the basic data
acquisition element being a cluster (5.7$\times$7.6 m$^2$),
made of 12 RPCs (2.8$\times$1.25 m$^2$ each). Each chamber is
read by 80 external strips of 6.75$\times$61.8 cm$^2$ (the spatial pixels),
logically organized in 10 independent pads of 55.6$\times$61.8
cm$^2$ which represent the time pixels of the detector. 
The read-out of 18360 pads and 146880 strips are the experimental output of the detector \cite{aielli06}.
The RPCs are operated in streamer mode by using a gas mixture (Ar 15\%, Isobutane 10\%, TetraFluoroEthane 75\%) for high altitude operation. The high voltage set at 7.2 kV ensures an overall efficiency of about 96\% \cite{aielli09c}.
The central carpet contains 130 clusters (hereafter, ARGO-130) and the
full detector is composed of 153 clusters for a total active
surface of $\sim$6700 m$^2$. 
A simple, yet powerful, electronic logic has been implemented to build an inclusive trigger. This logic is based on a time correlation between the pad signals depending on their relative distance. In this way, all the shower events giving a number of fired pads N$_{pad}\ge$ N$_{trig}$ in the central carpet in a time window of 420 ns generate the trigger.
This can work with high efficiency down to N$_{trig}$ = 20,
keeping the rate of random coincidences negligible.

The whole system, in smooth data taking since July 2006 firstly with ARGO-130, is in stable data taking with the full apparatus of 153 clusters since November 2007 with the trigger condition N$_{trig}$ = 20 and a duty cycle $\geq$85\%. The trigger rate is $\sim$3.6 kHz with a dead time of 4$\%$.
%

\begin{figure}[t!]
\begin{minipage}[t]{.47\linewidth}
\begin{center}
\epsfysize=4.5cm \hspace{0.5cm}
\epsfbox{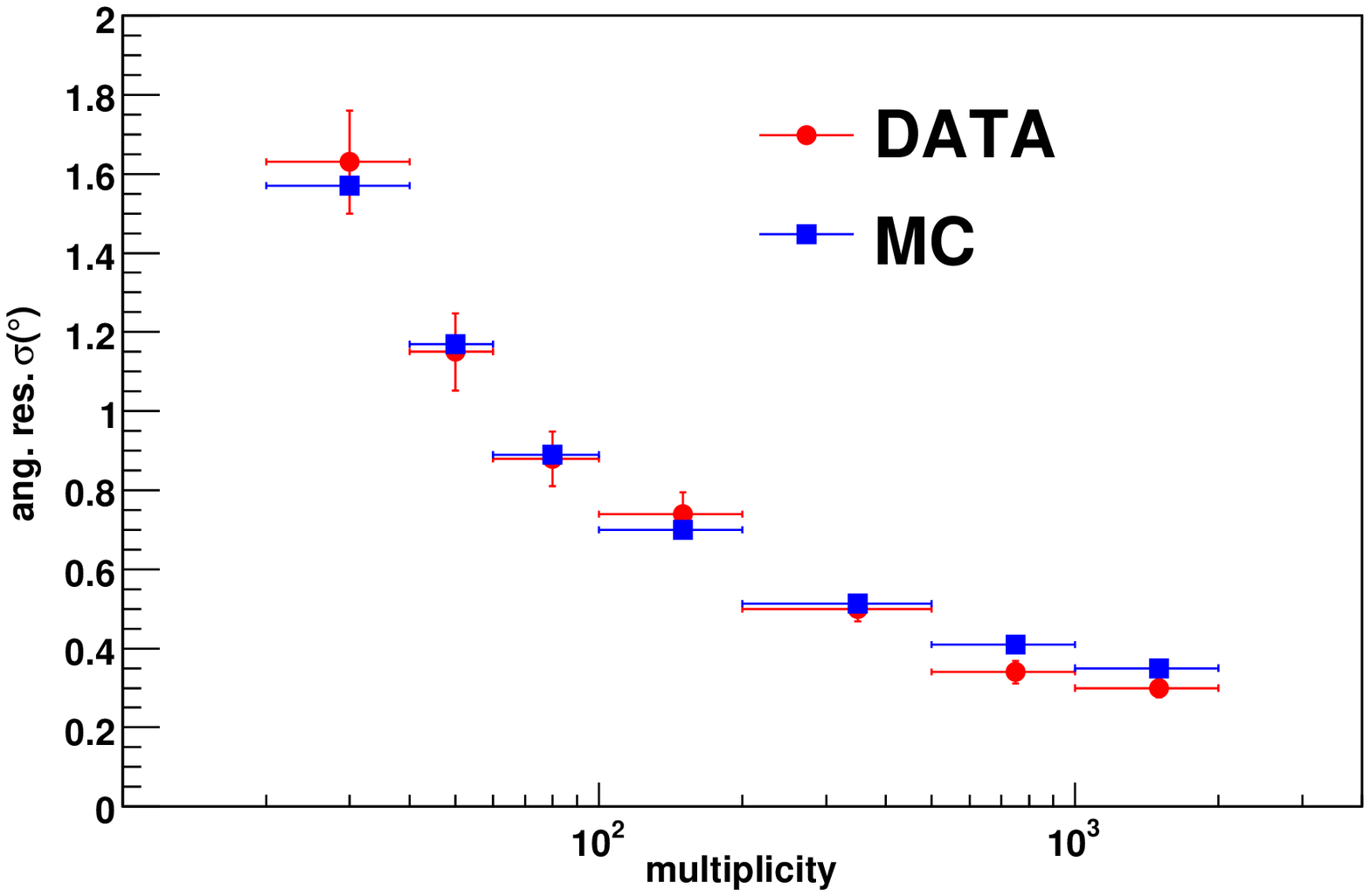} \vspace{-0.5cm}
\caption[h]{Measured angular resolution of the ARGO-YBJ detector
compared to expectations from MC simulation as a function of the
multiplicity. The used multiplicity bins are shown by the horizontal
errors.} 
\label{fig:angresol}
  \end{center}
\end{minipage}\hfill
\begin{minipage}[t]{.47\linewidth}
  \begin{center}
\epsfysize=4.5cm \hspace{0.5cm}
\epsfbox{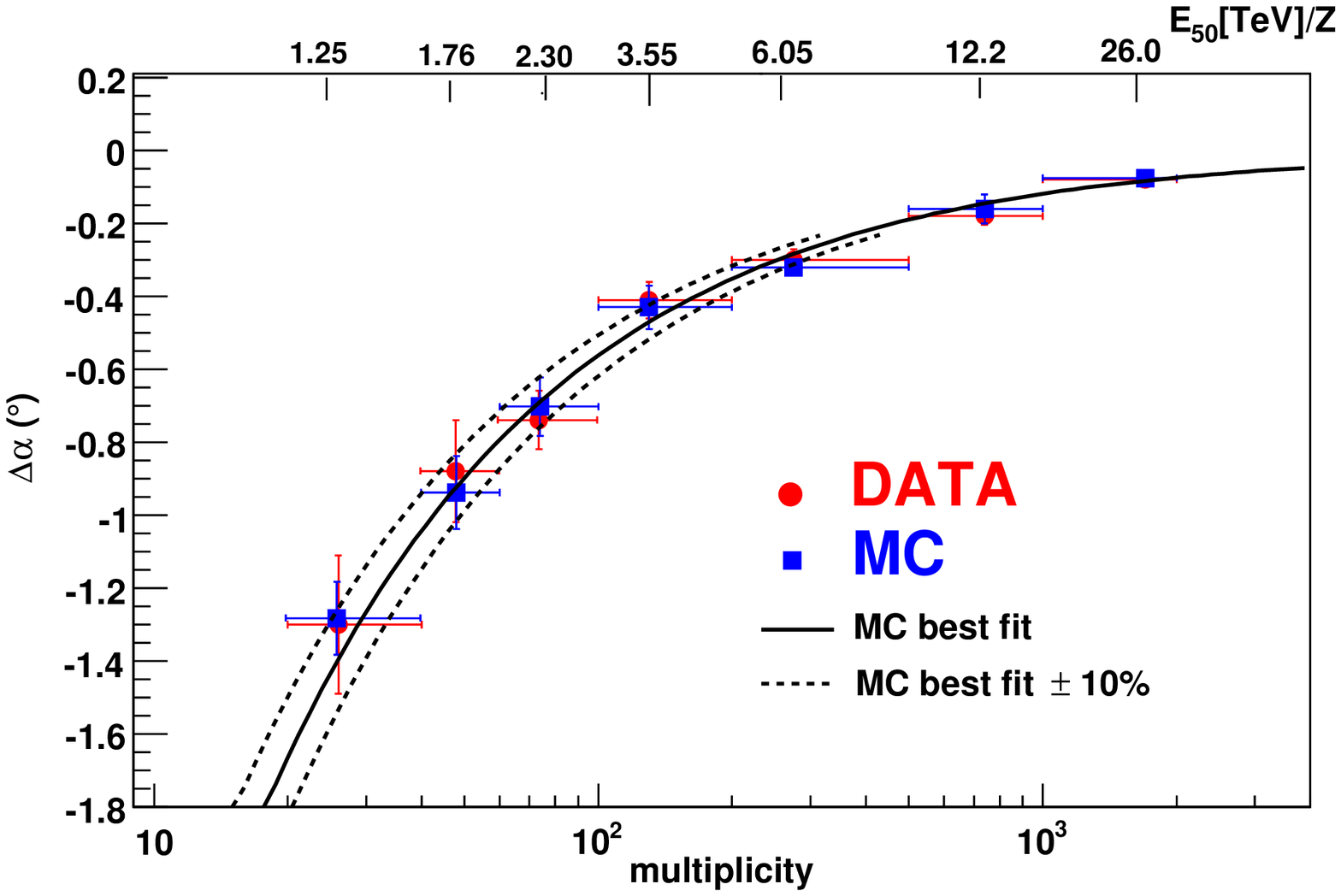} 
\caption[h]{Measured westward displacement of the Moon shadow as a function of multiplicity (red filled circles). The data are compared to MonteCarlo simulation (blue filled squares). The upper scale refers to the median energy of rigidity (TeV/Z) in each multiplicity bin.} 
\label{fig:DataMCEW}
  \end{center}
\end{minipage}\hfill
\end{figure}
%
\subsection{Detector performance}
The performance of the detector and the operation stability are continuously monitored by observing the Moon shadow, i.e., the deficit of CR detected in its direction. 
Indeed, the size of the deficit allows the measurement of the angular
resolution and its position allows the evaluation of the absolute pointing accuracy of the detector. In addition, positively charged particles are deflected towards East due to the geomagnetic field by an angle $\sim 1.6^{\circ}Z/E[TeV]$. Therefore, the observation of the displacement of the Moon provides a direct calibration of the relation between shower size and primary energy (Fig. \ref{fig:DataMCEW}).

ARGO-YBJ observes the Moon shadow with a sensitivity of $\sim$9 s.d. per month for events with a multiplicity N$_{pad} \geq$40 and zenith angle $\theta <$50$^{\circ}$, corresponding to a proton median energy E$_p\sim$1.8 TeV.
With all data from July 2006 to December 2009 (about 3200 hours on-source in total) we observed the CR Moon shadowing effect with a significance of about 55 s.d. \cite{vulcano10}.
The measured angular resolution is better than 0.5$^{\circ}$ for CR-induced showers with energies E $>$ 5 TeV (Fig. \ref{fig:angresol}) and the overall absolute pointing accuracy is $\sim$0.1$^{\circ}$.
The angular resolution stability is at a level of 10\% in the period November 2007 - March 2010 \cite{vulcano10}.
The energy scale error of ARGO-YBJ is estimated to be less than 18\% in the energy range 1 - 30 TeV/Z.
\section{The Crab Nebula}
We observed the Crab signal with a significance of 14 s.d. in about 800 days at a median energy of $\sim$2 TeV, without any event selection or gamma/hadron discrimination algorithm. 
In almost 3 years of data taking the observed flux was consistent with a steady emission, and the obtained differential energy spectrum in the 0.3-30 TeV range is dN/dE = (3.0$\pm$0.3)$\cdot$10$^{-11}\cdot$ (E/1 TeV)$^{-2.74\pm 0.14}$ cm$^{-2}$ s$^{-1}$ TeV$^{-1}$ in agreement with other measurements.

The distribution of the daily excess significances in the period November 2007 - October 2010 can be fitted with a Gaussian function with mean value 0.31$\pm$0.03 and r.m.s. = 0.99$\pm$0.02 ($\chi_{red}^2$ = 15/10), showing that the detector is performing correctly and that the Crab flux is stationary on one day scale over the investigated period.

\begin{figure}[t!]
\begin{minipage}[t]{.47\linewidth}
\begin{center}
\epsfysize=4.5cm \hspace{0.5cm}
\epsfbox{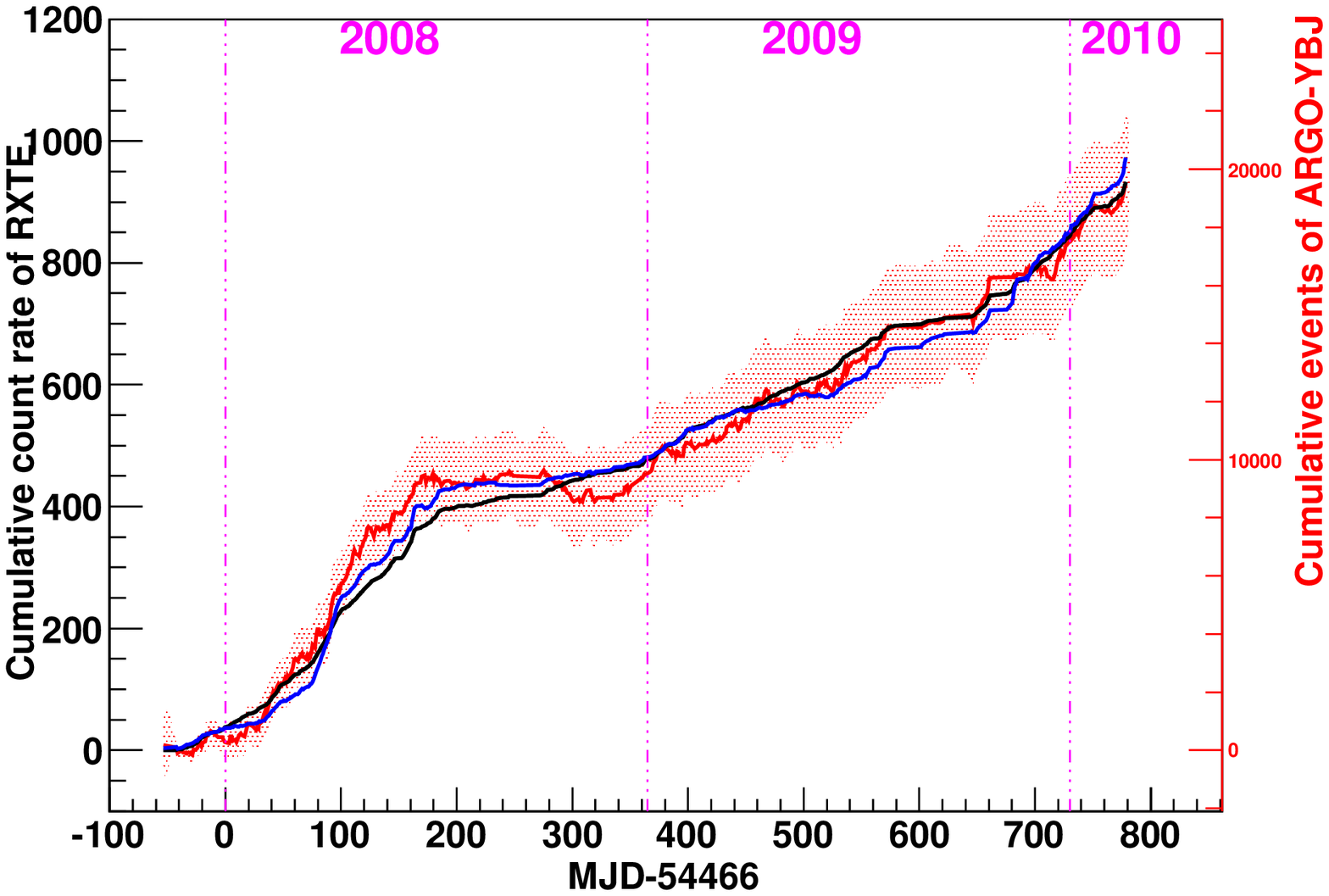} \vspace{-0.5cm}
\caption[h]{Cumulative light curve from Mrk421 measured by ARGO-YBJ compared with RXTE/ASM (black curve) and Swfit (blue curve) X-ray data.} 
\label{fig:mrk421_cumul}
  \end{center}
\end{minipage}\hfill
\begin{minipage}[t]{.47\linewidth}
  \begin{center}
\epsfysize=4.5cm \hspace{0.5cm}
\epsfbox{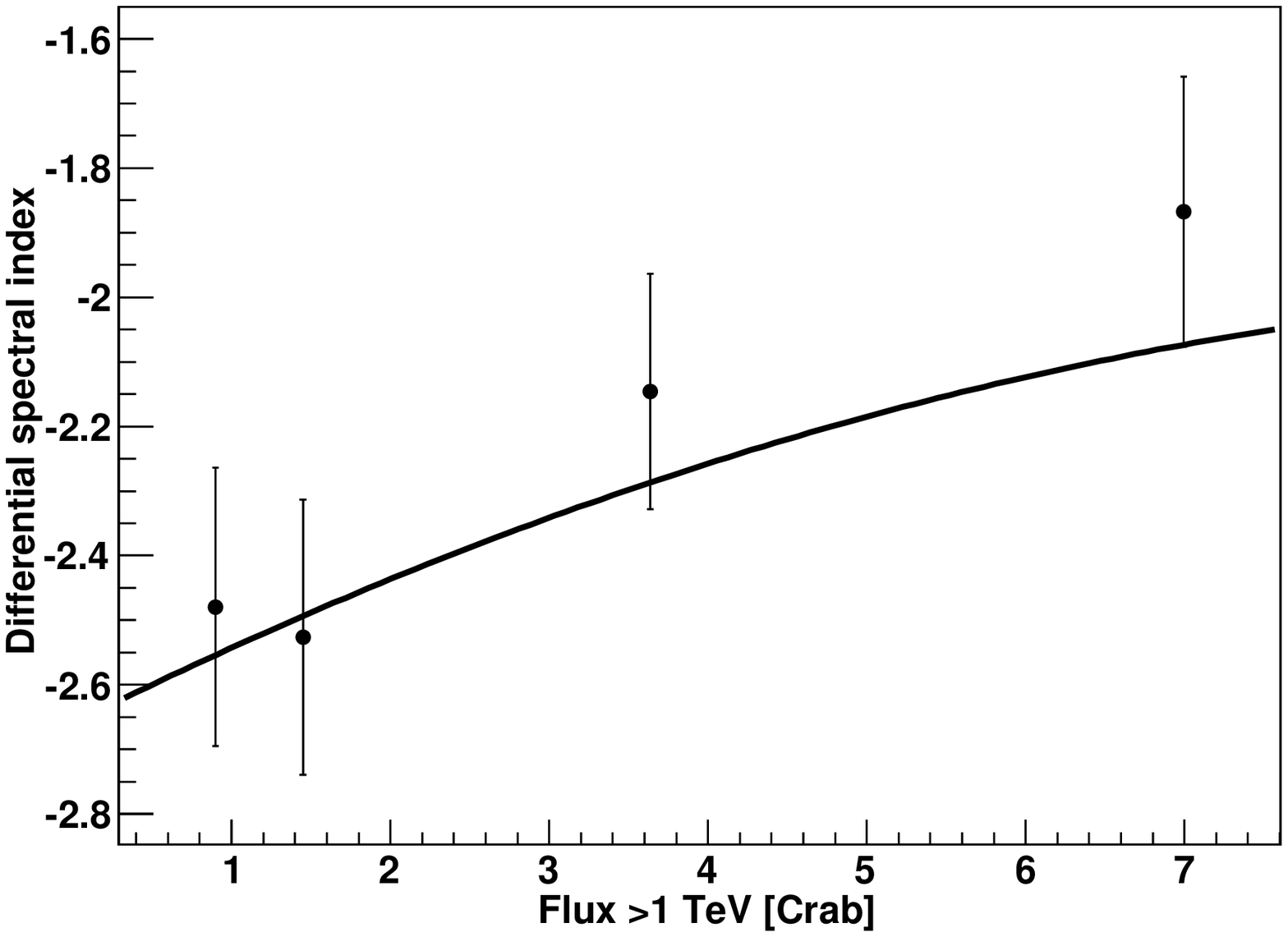} 
\caption[h]{Mrk421 spectral index vs. flux above 1 TeV in Crab units. The solid line is the function obtained in \cite{krenn02}.} 
\label{fig:mrk421-spindex}
  \end{center}
\end{minipage}\hfill
\end{figure}
%

\section{The blazar Markarian 421}

Mrk421 is one of the brightest blazars and the closest to us (z = 0.031),
characterized by a strong broadband flaring activity
with a variability time scale ranging from minutes to months.
A correlation of VHE gamma rays with X-rays has been observed
during single flaring episodes by different detectors
and can be interpreted in terms of the Synchrotron Self-Compton
(SSC) model. The Mrk421 high variability makes the long term multiwavelength
observation very important to constrain the emission mechanisms models.

Mrk421 was the first source detected by the ARGO-YBJ experiment in July 2006 when the detector started recording data with only the central carpet and was in commissioning phase.
ARGO-YBJ has monitored Mrk421 for more than 2 years, studying the correlation of the TeV flux with X-ray data. We observed this source with a total significance of about 12 s.d., averaging over quiet and active periods. Indeed, as it is well known, this AGN is characterized by a strong flaring activity both in X-rays and in TeV $\gamma$-rays. ARGO-YBJ detected different TeV flares in correlation with X-ray observations, as can be seen from Fig. \ref{fig:mrk421_cumul} where the ARGO-YBJ cumulative events per day is compared with the cumulative events per second of the RXTE/ASM and Swift satellites.
To get simultaneous data, 556 days have been selected for this analysis.
The X-ray/TeV correlation is quite evident over more than 2 years. The steepness of the curve gives the flux variation, therefore we observed an active period at the beginning of 2008 followed by a quiet phase. 

Cross-correlation functions were measured to quantify the degree of correlation and the phase difference (time lag) between the variations in X-ray and TeV bands, using the Discrete Correlation Function (DCF) \cite{edel98}. The DCF gives the linear correlation coefficient for two light curves as a function of the time lag between them.
For X-rays we consider two different bands (2-12 and 9-15 keV). 
The distributions peak around zero and the correlation coefficients around zero are $\sim$0.78. The peak positions, evaluated by fitting the distributions with a Gaussian function around the maximum, are -0.14$_{-0.85}^{+0.86}$ and -0.94$_{-1.07}^{+1.05}$ days for RXTE/ASM and Swift data, respectively.
Therefore, no significant time lag is found between X-ray data and TeV ARGO-YBJ observations.

In order to study the Spectral Energy Distribution (SED) of Mrk421 at different flux levels, both X-ray and TeV data have been divided into 4 bands based on the RXTE/ASM counting rate: 0-2, 2-3, 3-5, $>$5 cm$^{-2}$s$^{-1}$. For each band a SED with the average flux is calculated for different X-ray and $\gamma$-ray energies.
The RXTE/ASM detector is monitoring Mrk421 in three energy bands: 1.5-3, 3-5 and 5-12 keV. By assuming a power-law X-ray energy spectrum we calculated the spectral indices in the 4 flux bands: -2.02$\pm$0.08, -2.05$\pm$0.03, -2.15$\pm$0.03 and -2.43$\pm$0.04. A correlation is clearly recognized in the sense that higher fluxes are accompanied by harder energy spectra. This result, obtained by averaging the Mrk421 emission over about 2 years, extends similar conclusions obtained by averaging observations over much shorter periods \cite{rebil}.

With this information we studied the relation between TeV spectral index and flux. The TeV flux measured by ARGO-YBJ in the different bands ranges from 0.9 to about 7 Crab units.
As shown in Fig. \ref{fig:mrk421-spindex} the TeV spectral index hardens with increasing flux in agreement with the results obtained by the Whipple collaboration in 2002 (continuous line) \cite{krenn02}. 
This conclusion generalizes our result obtained with the analysis of the June 2008 flare \cite{aielli10}.

\begin{figure}[t!]
\begin{minipage}[t]{.47\linewidth}
\begin{center}
\epsfysize=4.8cm \hspace{0.5cm}
\epsfbox{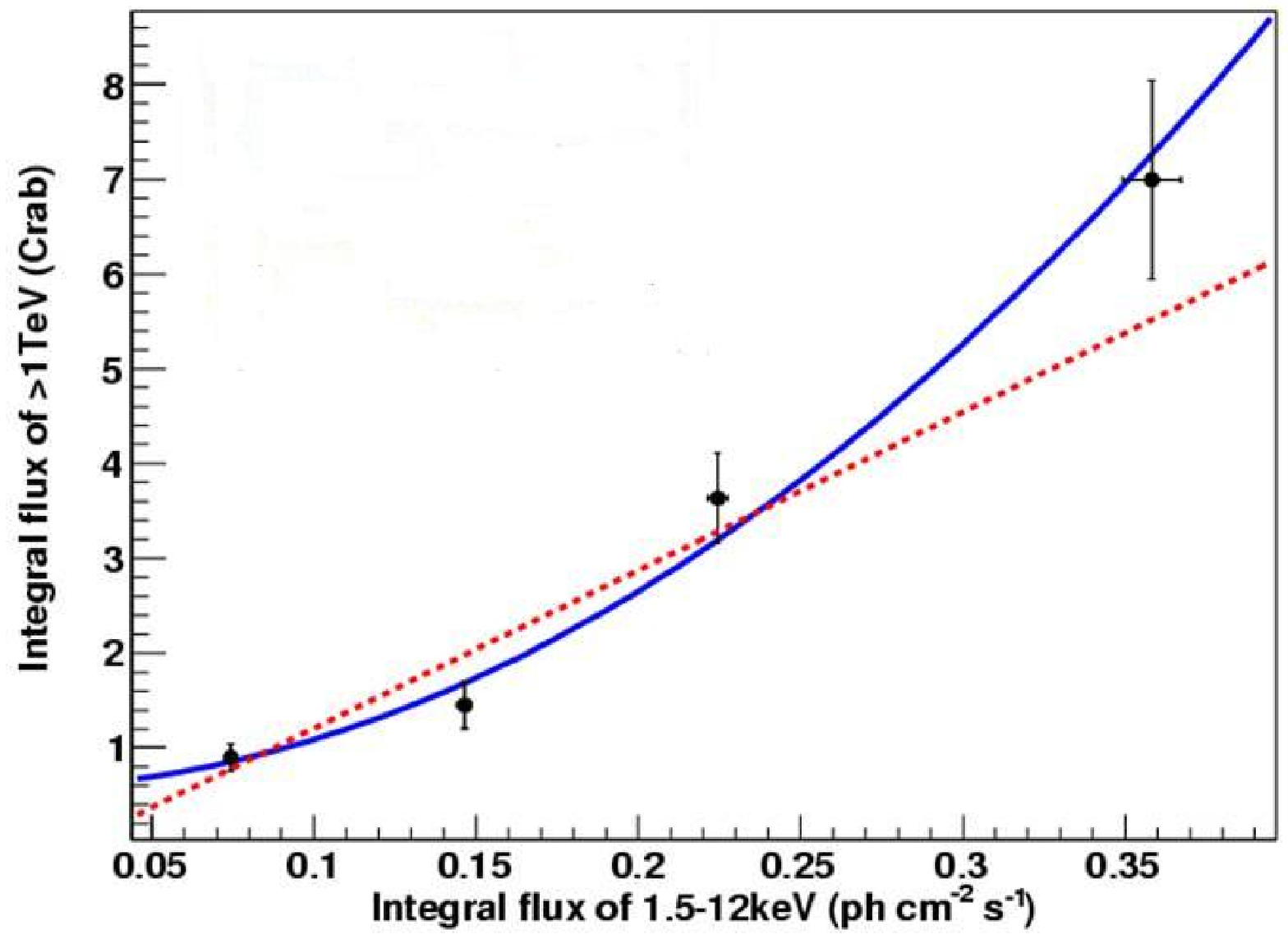} \vspace{-0.5cm}
\caption[h]{ARGO-YBJ $>$1 TeV flux vs. RXTE/ASM X-ray flux. 
Solid line: quadratic best-fit ($\chi^2$/dof= 1.9/2). Dotted line: linear best-fit ($\chi^2$/dof= 7.7/2).} 
\label{fig:mrk421-quadratic}
  \end{center}
\end{minipage}\hfill
\begin{minipage}[t]{.47\linewidth}
\vspace{-0.5cm}
  \begin{center}
\epsfysize=6.cm \hspace{0.5cm}
\epsfbox{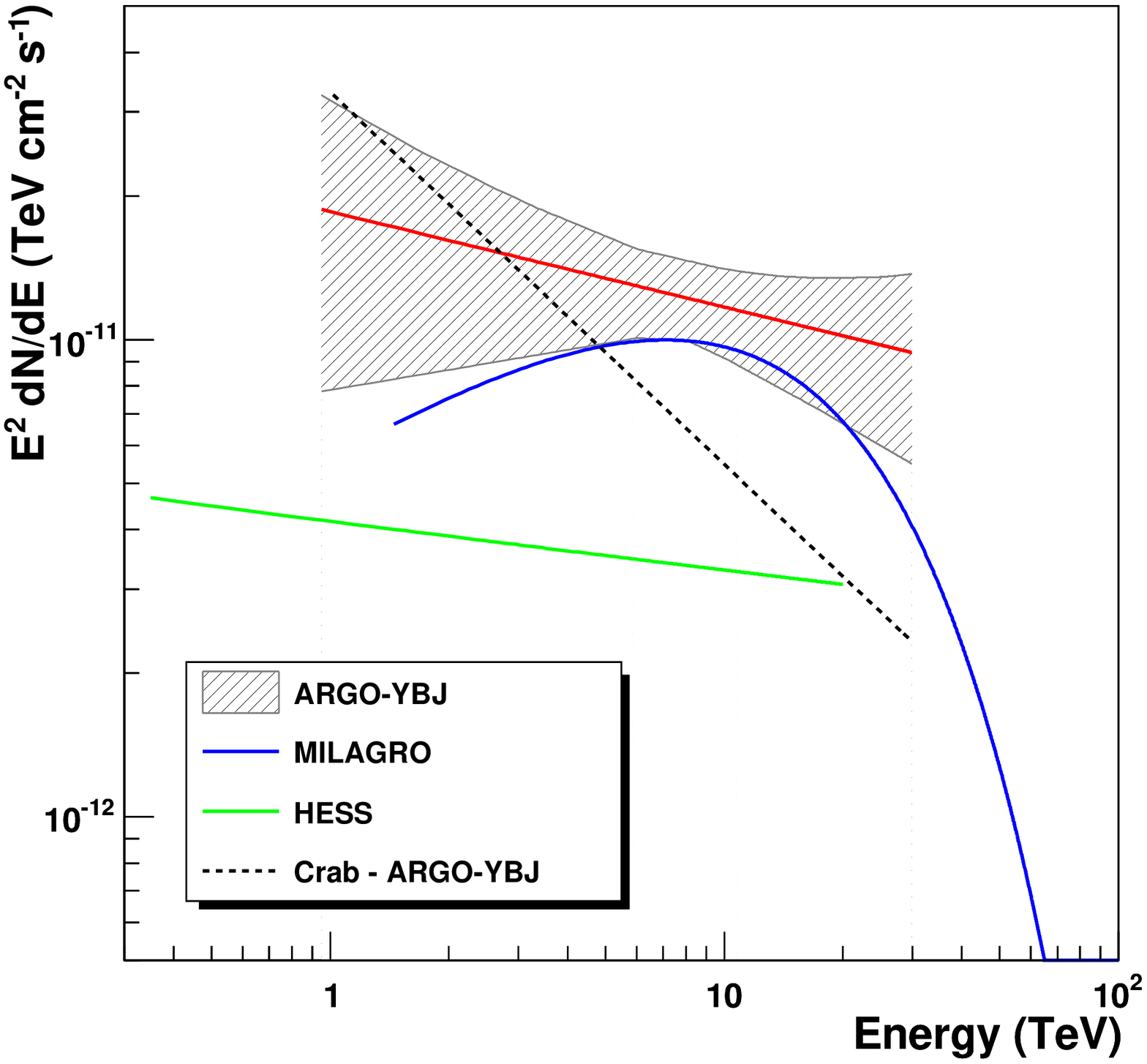} \vspace{-0.5cm}
\caption[h]{Differential flux from MGRO J1908+06 measured by ARGO-YBJ (red line) compared with MILAGRO and HESS spectra. The shaded band represents 1 s.d. error. The black dashed line represents the Crab Nebula flux.} 
\label{fig:ene_spe}
  \end{center}
\end{minipage}\hfill
\end{figure}

We investigated also the correlation between RXTE/ASM X-ray and ARGO-YBJ TeV fluxes (Fig. \ref{fig:mrk421-quadratic}). A positive correlation is clearly observed and the X-ray/TeV relation seems to be quadratic rather than linear, in agreement with the results of \cite{fossati}.

%

\section{MGRO J1908+06}

The gamma ray source MGRO J1908+06 was discovered by the MILAGRO 
\cite{mila07} at a median energy of $\sim$ 20 TeV and recently associated with the Fermi pulsar 0FGL J1907.5+0602 \cite{mila09}.
The data were consistent both with a point source and with an extended
source of diameter $<$ 2.6$^{\circ}$.
HESS confirmed the discovery with the detection of the extended source HESS J1908+063 \cite{hess09} at energies above 300 GeV, positionally consistent with the MILAGRO source. The extension of the source was extimated $\sigma_{ext}$ = 0.34$^{\circ}$$_{-0.03}^{+0.04}$. 
The Milagro and HESS energy spectra are in disagreement at a level of 2-3 s.d., being the Milagro result about a factor 3 higher at 10 TeV \cite{mila09b}.

ARGO-YBJ observed a TeV emission from MGRO J1908+06 with a significance of 6 s.d. in about 730 days.
At the ARGO-YBJ site, MGRO J1908+06 culminates at the relatively low zenith angle of 24$^{\circ}$ and is visible for 5.4 hours per day with a zenith angle less than 45$^{\circ}$.
By assuming a 2D Gaussian source shape with r.m.s = $\sigma_{ext}$, 
independent of the gamma ray energy, we fitted the event distribution as a funcion of the distance from the center (set to the Fermi pulsar position)
and, taking into account the PSF, 
we found $\sigma_{ext}$=0.48$^{\circ}$$_{-0.28}^{+0.26}$, a value larger but consistent with the HESS measurement.

The best fit power law spectrum is: dN/dE = 3.6$^{+0.7}_{-0.8}$ $\times$ 10$^{-13}$ (E/6 TeV)$^{-2.20^{+0.34}_{-0.29}}$ photons cm$^{-2}$ s$^{-1}$ TeV$^{-1}$ (the errors on the parameters are statistical) (Fig. \ref{fig:ene_spe}).
The systematic errors are mainly related to the background evaluation and to 
the uncertainty in the absolute energy scale. According to our estimate,
they globally affect the quoted fluxes for $\leq$ 30$\%$.
Given the flat slope, above a few TeV the MGROJ1908+06 flux becomes larger than the Crab Nebula one.
A significant disagreement appears between the ARGO-YBJ and HESS spectra. 
Concerning ARGO-YBJ and MILAGRO, considering the large errors in the flux measurements (the error on the MILAGRO flux is $\sim$30$\%$ at 10 TeV 
and larger at lower energies), the apparent disagreement could be likely due to statistical fluctuations. In the limit of the statistical accuracy of this result, our data supports the MILAGRO measurement of a flux significantly larger than that measured by HESS.
One possible cause of this discrepancy is that ARGO-YBJ and MILAGRO integrate the signal over a solid angle larger than the HESS one, and likely detect more of the diffuse lateral tail of the extended source. A further interesting possibility is the variability of the source, since the HESS data consist of 27 hours of observations in 2005-2007, before our measurements.

\end{document}